SSC26-FT-75

# The Billion Dollar Surprise: How Solar Cycle 25 Cut Satellite Lifetimes in LEO

Scott Shambaugh
Leonid Space
leonidspace.com/contact

## ABSTRACT

Solar Cycle 25 has run far stronger than the 2019 consensus forecast issued by the NOAA/NASA/ISES prediction panel, with densities in low Earth orbit from 2022-2026 holding at 2-3x the predicted levels. The cumulative drag impulse experienced by LEO satellites, which is the driving metric for propellant budgets and mission lifetimes, reached 5-6 standard deviations beyond the forecast's stated uncertainty. This means that even operators who designed conservatively against the two-sigma worst case fell short of their drag budgets. This paper quantifies a lower bound on the economic cost of that misprediction.

Starting from the 13,704 payloads on-orbit below 800 km during 2022-2026, we screen to the 1,597 payloads which we were able to validate with high confidence to be both operational and in ballistic freefall without propulsive capabilities, representing 4.8% of all satellite value in LEO. We estimate each satellite's ballistic coefficient and propagate its trajectory under the forecasted atmosphere versus the observed one. A probabilistic cost model assigns each satellite an annualized mission cost based on direct costs (amortized capital costs plus annual operations), stratified by size class, with bespoke estimates for high value missions. Survival and forward cost discounting is applied at a modal 11% per year. We combine the differences in lifetime with a Monte Carlo sampling of the cost model to estimate the total dollar impact of the solar cycle misprediction on satellite operators.

Against the forecast's two-sigma upper bound which we consider to be a standard engineering design target, these satellites lost 688 cumulative mission years valued at $0.88 billion. Against the nominal forecast, these satellites lost 2,472 mission years worth $2.77 billion. NASA's Swift mission is used as a case study to show how its expected direct value loss is comparable to the price of its pending reboost rescue mission. These estimates are deliberate lower bounds which exclude propulsive satellites (the remaining 95.2% of LEO value), revenue above direct cost, and downstream economic impact. The results give a quantitative case for the value of accurate decadal-scale space weather forecasting, and show that well-calibrated uncertainty bounds are as valuable to a satellite operator end-user as the accuracy of the central prediction itself.

## INTRODUCTION

Satellites in low Earth orbit (LEO) live and die by atmospheric drag. The drag a satellite experiences is driven by neutral density in the thermosphere, which can vary by an order of magnitude between solar minimum and solar maximum, as extreme ultraviolet (EUV) radiation heats and expands the upper atmosphere throughout the 11-year solar cycle. Because spacecraft must be designed years before launch and operated for years afterwards, the entire economics of a LEO mission – orbit selection, propellant budgets, design lifetime, revenue forecasting and financing, production and replenishment cadence – rests on a prediction of solar activity a decade or more into the future.

For Solar Cycle 25, the consensus prediction was the 2019 forecast put out by an international NOAA/NASA/ISES panel, which called for a weak cycle in line with the quiet Cycle 24.[1] Cycle 25 did not cooperate. Instead, it has peaked sooner and stronger than the forecast. Densities in LEO from 2022-2026 consistently ran 2-3x higher than expected, and the cumulative delta-V from drag ran five to six standard deviations beyond the Gaussian interpretation of its stated uncertainty. The public announcements by a few satellite operators are the visible consequences of a drag environment that has affected the entire industry. This paper quantifies that impact by screening the full LEO catalog to 1,597 operational satellites validated to be in ballistic freefall, propagating each under the forecast versus the observed environment, and pricing the difference. The result is $2.77 billion in lost mission life against the nominal forecast, and $0.88 billion against its two-sigma worst case. This grounds the value of accurate decadal-scale solar cycle forecasts, and underscores the importance of well-calibrated uncertainty bounds.

## CHARACTERIZING THE MISPREDICTION

### *SWPC 2019 Panel Prediction*

Since 1979, an international panel of solar dynamics experts has convened during solar minimum to forecast the strength and timing of the upcoming 11-year cycle. The panel to predict Solar Cycle 25 met in 2019 and was cosponsored by NOAA, NASA, and ISES. We will refer to its prediction as the NOAA Space Weather Prediction Center (SWPC) 2019 Panel prediction. Miesch (2025) provides an overview of the panel proceedings.1 While other solar cycle forecasts exist, the international consensus panel prediction has served as the gold standard for satellite operator end-users.

### *Drag, Density, and Cumulative Dynamic Pressure*

The solar cycle is typically forecasted via the metrics of sunspot number (SSN) or solar flux in the 10.7 cm radio band (F10.7), which are proxies for the EUV radiation that drives thermosphere density over long timescales.

From a LEO satellite operator's perspective, SSN and F10.7 are not directly useful values. Solar flux needs to be converted to atmospheric density via an atmosphere model, so that they can model the forces and torques acting on the spacecraft. For some use cases such as short-term orbit propagation and control system sizing, engineers need the point densities along the orbit path. For satellite mission planning and propulsion system sizing, they need the cumulative delta-V from drag ($dV_{drag}$) over the entire lifetime of the satellite. Of these, $dV_{drag}$ is a harder binding constraint – system sizing against point densities is usually driven by end-of-mission control authority at the high-density low altitudes shortly before reentry, which is a days-to-weeks long environment and outside of normal operations. $dV_{drag}$ on the other hand, is cumulative during the extent of the mission, at normal operational altitude.

$dV_{drag}$ can be parameterized as the product of the ballistic coefficient of a spacecraft $C_B$, and the cumulative dynamic pressure $Q_C$ it sees in its orbit. This allows for separation of spacecraft-specific properties from environmental properties.

$$dV_{drag} = C_B Q_C \tag{1}$$

The cumulative dynamic pressure over a time period is measured in units of total impulse from drag per unit area, N-s/m^2 and is a function of the orbit-average density $\rho$ and object velocity relative to the airstream $v_{rel}$. $C_B$ is a constant function of drag coefficient $C_D$, frontal area $A$, and satellite mass $m$.

$$Q_C = \int_{t_0}^{t_1} \frac{1}{2} \rho v_{rel}^2 \, dt \tag{2}$$

$$C_B = \frac{C_D A}{m} \tag{3}$$

Figure 1 shows Solar Cycle 25 flux, density, and $Q_C$ from the perspective of a satellite in a representative circular 500 km altitude, 53 deg inclination orbit, assuming a NRLMSISE-00 atmosphere.2 The raw values for different orbits will vary by altitude and inclination, but the pseudo-exponential scaling of the atmosphere means that the ratios of these curves remains qualitatively the same throughout LEO.

### *Panel Prediction Uncertainty*

The SWPC 2019 panel gave its prediction as a peak SSN of 115 ± 10, with a curve of SSN over time and little guidance on how to interpret that uncertainty. Using the conversion between SSN and F10.7 from Clette (2021), this equates to a flux peak of 136 ± 8 SFU.3 We assume that this represents one standard deviation for a Gaussian distribution. To avoid unrealistic over-dispersion during the more predictable solar minimum, we apply 100% of this dispersion at solar max, and scale it linearly to 50% (± 4 SFU) at the minimum tails of Solar Cycle 25.

This interpretation is conservative for the design-case comparisons used later in the paper. If operators instead interpreted the panel's ±10 SSN as a two-sigma interval, or as a maximum expert range, then the corresponding uncertainty bands would be narrower than those used here. Their resulting drag budgets would be lower, and the estimated losses relative to those design cases would increase, falling between the nominal and 2σ results we present.

### *Impact on Satellites*

Satellite operators must design their spacecraft years before launch, for a vehicle design that may be produced and launched for several years, each of which will be flying for 5, 10, or more years. Accurate modeling of the space environment one to two decades into the future is therefore necessary to ensure the spacecraft is built such that it can perform its mission.

The two main knobs to turn for determining a satellite's mission life are propulsion system sizing and orbit selection, both of which are highly mass-sensitive trades which must be optimized against other system budgets. Precise forecasting of solar flux on the decadal timescale is not possible, so an engineer must design using uncertainty bands so they can survive realistic worst-case scenarios. A common metric is to design to cover 95% of plausible scenarios – protecting for "two sigma" uncertainties under a Gaussian distribution assumption. To this end, well-quantified uncertainties can be more important to space weather end users than accuracy of the central forecast.

Figure 1 shows that the cumulative dynamic pressure has risen to 5-6σ levels over the course of Solar Cycle 25, with density spikes to higher sigma-levels than that. To put it another way, by looking at the plot of cumulative dynamic pressure we see that a satellite launched in 2022 that was designed with enough fuel to handle the 2σ case would be short by 1/3 of its total fuel capacity by 2026.

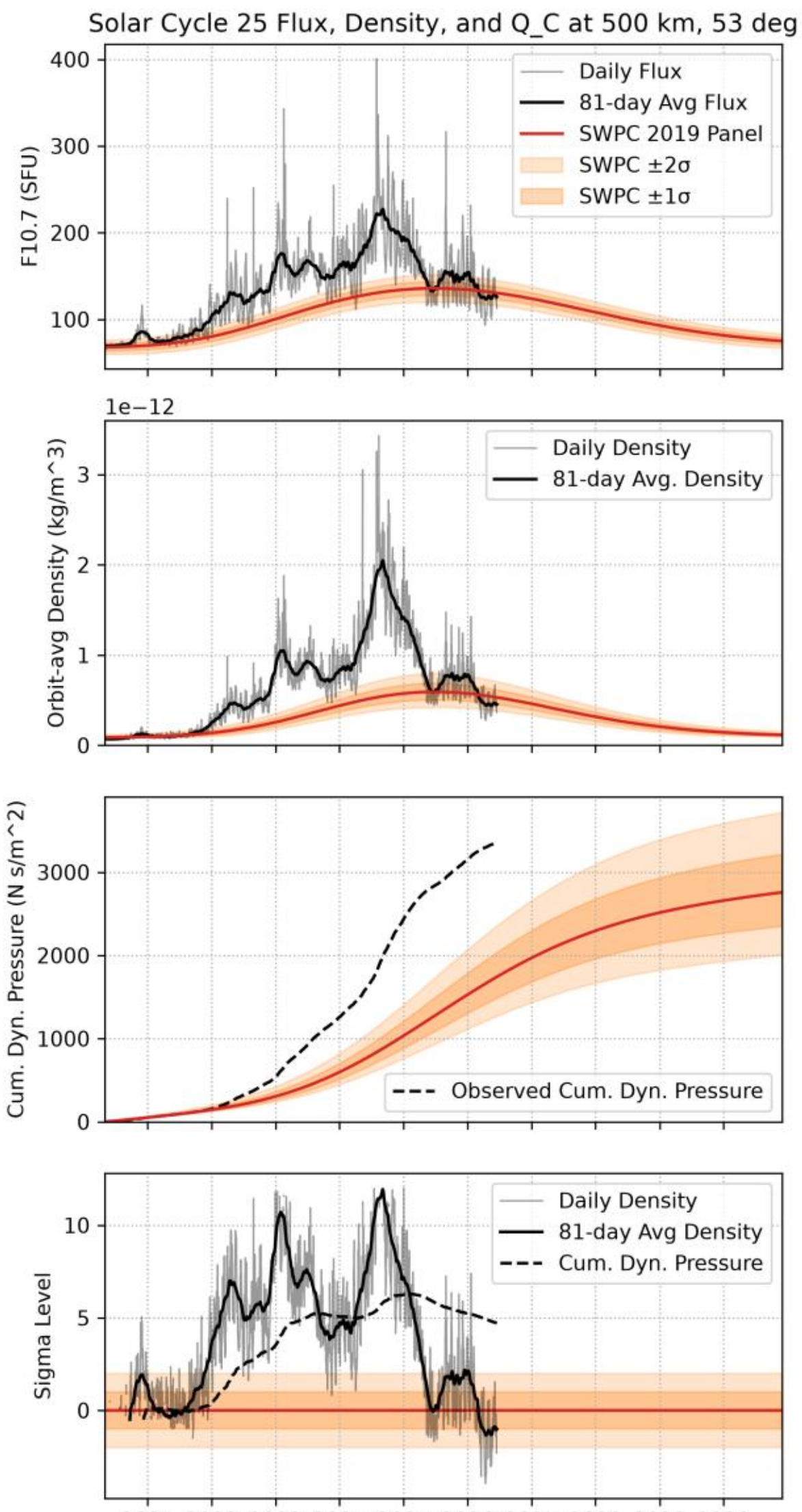

**Figure 1: Observed versus predicted F10.7, density, cumulative dynamic pressure, and sigma levels of these during Solar Cycle 25, for a representative 500 km, 53 deg orbit.**

## ECONOMIC IMPACT

Several high-profile announcements have underscored the extent to which satellite operators were surprised by and unprepared for the higher-than-expected drag:

- Capella Space reported losing several spacecraft in 2023 due to cumulative drag, with their expected design life of 3 years falling to 9 months given revised lifetime estimates.4
- NASA discovered in the middle of 2025 that the flagship Neil Gehrels Swift Observatory was due to deorbit in mid-2026, prompting the commissioning of a $30M rescue mission to rendezvous and push Swift to a higher altitude.5,6
- Planet implemented a "low drag" mode of flight to increase the lifespan of its non-propulsive Dove satellites.7

However, these few cases are just what has been publicly announced. We believe this is just the tip of the iceberg for impact from drag this solar cycle. Every operator in LEO experienced the same environment – whether their lack of disclosure represents a lack of impact or a desire to keep underperformance quiet will vary from mission to mission.

### *Economic Impact Assessment*

We quantify the economic impact of Solar Cycle 25's drag environment as compared to the expectations under the SWPC 2019 Panel prediction, across the entire industry. We estimate $C_B$ from on-orbit two-line element (TLE) data provided by the US Space Force through space-track.org, for the reference set of all operational payloads in LEO during Solar Cycle 25 (2022-2026) which are producing ongoing economic value through their operations.8 We propagate those satellites under the SWPC counterfactual and compare how much longer they would have lasted as compared to the observed space weather environment. We then construct a cost model of the annualized mission cost (AMC) for each satellite to estimate the value of a marginal year of operations. The economic impact on a single satellite is that marginal value multiplied by the difference in lifetime, and that impact is summed over the whole set of reference satellites.

$$Impact = \sum_{sats} AMC * \left(Life_{SWPC} - Life_{Obs/Est}\right) \quad (4)$$

## HANDLING THE MAJOR UNKNOWNS

How many years of useful life were erased by the higher-than-expected solar cycle? What was the value of that lifetime to satellite operators? There are significant unknowns in answering both of these questions. Satellite production, launch, and operational costs are closely held proprietary information, as are the expected design life, propulsive capabilities, and orbit control plans for any given on-orbit asset. Whether a satellite is still operational and how much revenue it generates per month is not typically shared, and few assessments of downstream customer value have been performed. Our analysis therefore takes a conservative approach to estimating impact.

Firstly, we exclude all satellites which are known to have propulsive capabilities, or have demonstrated altitude control on-orbit. This is not to say that satellites without propulsion systems on board were not also negatively affected – their fuel reserves will have drained at a higher rate to counteract drag and maintain altitude, and depending on design margins and the sensitivity of the mission to precision orbit control these may be just as negatively affected as the satellites in ballistic freefall. But propulsive satellites at least have the potential to have carried enough onboard fuel to counteract SC25's drag without an overall impact on the mission. So we exclude them. This excludes the satellites in the Starlink, OneWeb, and Leo (Kuiper) megaconstellations, which make up the large majority of on-orbit assets but are known to have propulsion on-board.

For satellites in ballistic freefall, it is unknown what their planned mission duration was, and what design margin on this mission lifetime was built into orbit selection and vehicle design. So we assume that operators designed their vehicles to survive the 95% confidence interval worst case space weather prediction (the SWPC forecast dispersed 2$\sigma$ high), and compare against the drag in that scenario. We also report out the difference against the nominal space weather forecast, but do not believe this to reflect normal design practices.

How many of these satellites in ballistic freefall are still operational, versus already having failed on-orbit? We apply a two-tiered filter. Firstly, large mass payloads typically represent high-value national security or civil space missions, for which operation status is publicly available. When this information was available, we manually excluded non-operational satellites. For the rest, Celestrak lists the operational status for a number of LEO payloads.9 This has limited coverage of the catalog, however we can filter by taking this as a base rate of on-orbit failures and apply that as a knock-down factor on the value for satellites where operational status is uncertain.

That value of an extra year on orbit is also uncertain. For large high-value satellites, we produce custom estimates of amortized mission cost. For the rest, we stratify the catalog by mass class and build a cost model that incorporates reasonable estimates of what it costs to build, launch, and operate satellites of a given size. That value is divided by estimated expected lifetime to give the yearly replacement value of any given satellite. Downstream economic impact is ignored, and only these direct costs are modeled. The cost model is built using probability distributions rather than point estimates to reflect the underlying uncertainty, and run through a Monte-Carlo simulation to find expected impact.

We expect this approach to produce a realistic conservative lower bound on total economic impact. Table 1 summarizes our handling of these major unknowns. In particular, modeling of the drag impact on propulsive satellites, surplus revenue beyond direct costs, and the economic impact on downstream users would all tend to drive true impact higher.

**Table 1: Major Unknowns and Their Modeling**

| Unknown | Modeling Approach |
| --- | --- |
| Existence and capabilities of on-board propulsion | Restrict analysis to satellites in ballistic freefall |
| Planned mission duration and design margin | Compare to 95% CI high drag forecast |
| Operational status | Direct screening and Celestrak base-rate knockdown |
| Cost to build, launch, and operate satellites | Estimate based on size class and bespoke research |
| Downstream economic impact | Ignore, direct costs only |

## SATELLITE SELECTION

We applied a multi-stage selection filter to arrive at the set of operational non-propulsive payloads. These criteria were developed iteratively, guided by the data. Thresholds were tuned and edge cases manually inspected until the surviving population consistently represented satellites in ballistic freefall.

### *Initial Screening to LEO Payloads*

We first collected all objects marked as PAYLOAD in the space-track.org "SATCAT" catalog, and restricted this list to the objects that were in orbit between 2022-01-01 and 2025-12-31 for at least 30 days. The first TLE of the object after launch was checked to ensure that both perigee and apogee altitudes were between 80 – 800 km to restrict the analysis to the drag-dominated regions of LEO, and the last TLE was restricted to below 800 km.

Figure 2 shows the dynamics of the population of payloads in LEO. Long-duration missions hold steady at their altitude bands, while megaconstellation satellites insert at low altitudes and use onboard propulsion to raise to their operational orbits. A steady rain of satellites falls back to Earth throughout the solar cycle, until they become lost to tracking and stop reporting TLEs beneath ~200 km.

### *Megaconstellation Exclusion*

Satellites in the Eutelsat OneWeb, SpaceX Starlink and Amazon Leo (Kuiper) constellations were manually excluded due to known propulsive capabilities. OneWeb had already been largely excluded due to those satellites operating at 1200 km.

### *Inert Objects, Orbital Transfer Vehicles (OTVs), and Human Spaceflight Exclusions*

Inert objects with no economic value such as dummy payloads, mass simulators, passive deployment structures, adapter rings, and payload deployers were excluded via comparison against manually identified naming patterns. Orbital transfer vehicles and kick stages (some of which were listed as PAYLOAD rather than ROCKET_BODY) were screened out. We also excluded all known human spaceflight objects, including crewed vehicles, cargo transfer vehicles, and space stations.

### *Altitude Raise Detection*

Any satellites where the semi-major axis altitude rose more than 2 km at the last TLEs as compared to the TLE 30 days prior were excluded, as evidence of propulsion. This 2 km allowance accommodates noise on TLEs.

### *VLEO Screening*

Any satellite with an initial semi-major axis altitude below 350 km was screened out due to either propulsion being required to maintain altitude, or insufficient tracking data over the short remaining lifetime.

### *Long-life Low Mass Satellites*

Mass categories for each satellite from ESA's DISCOSweb catalogue were collected, and all satellites with a mass less than 200 kg (Micro, Nano, Pico, and Femto) were checked to see if they had already been on orbit for more than 10 years.10,11 These were excluded as prima facie evidence of either propulsion, or lifetime beyond design expectations and likely non-operation.

## ORBIT PROPAGATION AND SCREENING

### *Ballistic Coefficient Calculations and Plausibility Screening*

Ballistic coefficients ($C_B$) for all remaining satellites were estimated using Leonid's proprietary pipeline as described in Shambaugh (2026).12 This assumes a NRLMSISE-00 atmosphere driven by historical F10.7 and Ap geomagnetic space weather indices, and uses energy dissipation rates along with noise filtering to estimate mission-average constant "cannonball" $C_B$ values. Any satellite with a $C_B$ less than 0.001 m^2/kg

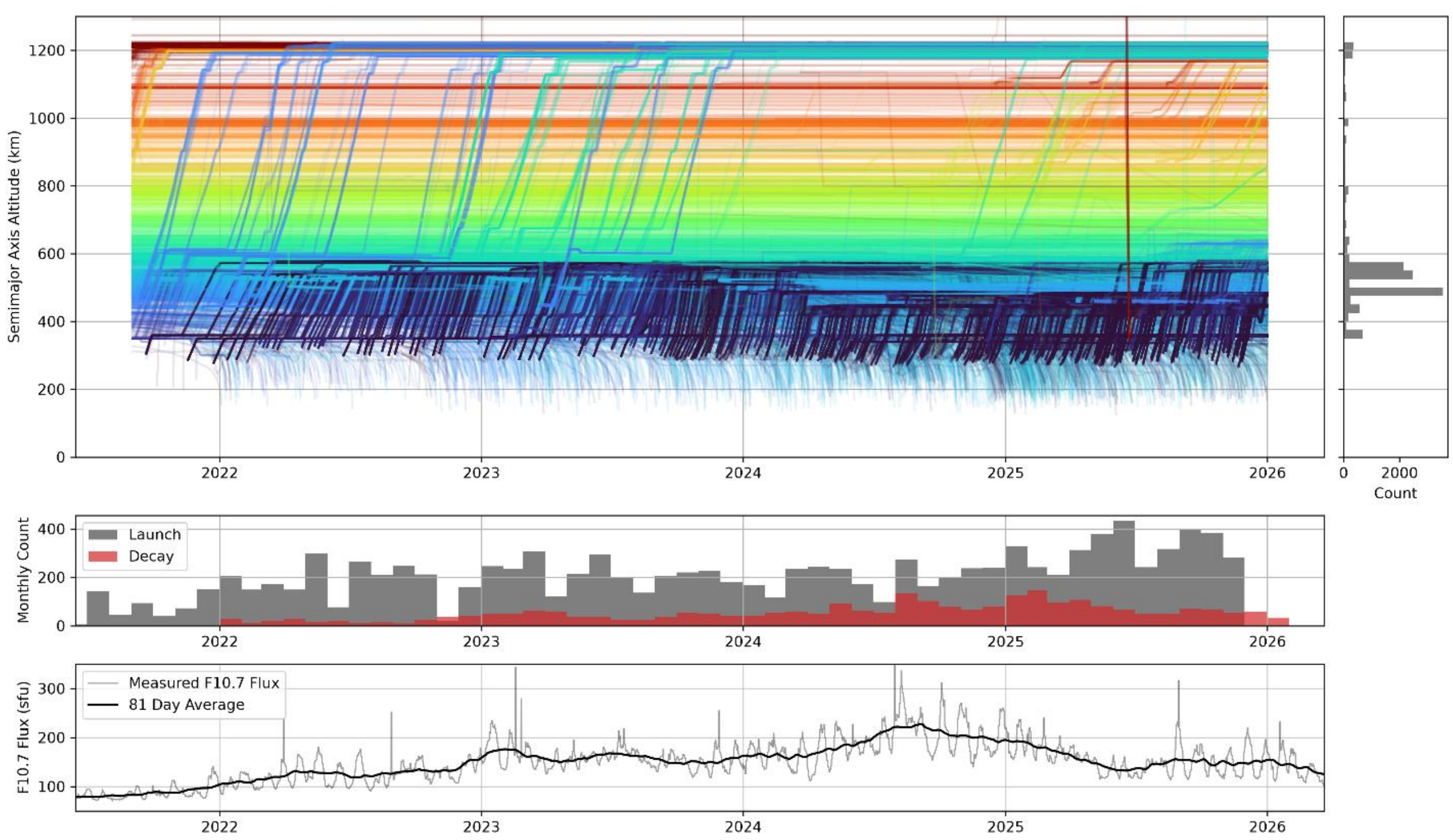

**Figure 2: LEO PAYLOAD satellite population from 2022-2026. Top plot shows the semi-major axis altitude, colored by starting altitude. Middle plot shows the number of objects launched and decayed each month. Bottom plot shows F10.7 solar flux over this time period.**

was deemed to be implausibly low (indicative of active orbit control or raising) and excluded.

### *Propagation Approach*

We used a custom propagator described in Shambaugh (2026).12 This uses a modified King-Hele approach whereby mean elements are propagated using orbit-averaged NRLMSISE-00 atmospheric densities in a co-rotating atmosphere.13 Use of the same model to perform ballistic coefficient estimation and propagation absorbs the potential errors from bias offsets between different atmosphere models.

For SWPC predictions, the atmosphere model is fed by F10.7 solar flux values from the SWPC 2019 panel forecast with an assumed constant geomagnetic index of Ap=14, until the end of the prediction time horizon in 2030. Note that storm events are implicitly covered under this average Ap value – the higher density during storm events only lasts on the order of hours, and does not integrate up into noticeable impact on total $dV_{drag}$. For baseline predictions, the measured F10.7 and Ap values provided by GFZ are used up until 2026-06-01, at which point it switches to Marshall Spaceflight Center's (MSFC's) MSAFE forecasts.14,15,16,17 The MSAFE forecast model is not as widely used, but tracks the historical mean for future cycles and is updated monthly to better fit recent observed flux. It is built from a statistical collection of past cycles and bounds all past observations well.

Shambaugh (2026) describes the validation campaign that was performed to quantify accuracy of Leonid Space's $C_B$ estimation and propagation pipeline. Over a one-year forecasting horizon on a test set of 934 satellites with an estimated ballistic coefficient and known space weather, it produces a median CRPS error of 18.6 days (5.1%). This measured all process noise and errors introduced from missing higher-order atmosphere model and orbit propagation dynamics, and all variation above this is attributable to errors in solar flux predictions.

### *Solar Cycle 26*

For satellites still in orbit, the years of life lost compared to the panel prediction counterfactual is not yet known. This timing difference will be driven by how strong Solar Cycle 26 turns out to be, and the effect of SC25 will have been to set up the entry conditions to SC26. A strong cycle will act as a "wall" that will bring satellites down faster and compress the time between the baseline and panel counterfactual deorbit events, in essence weakening the effect of SC25 misprediction. A weak cycle will spread out this time, and result in a larger differential in lost years of life. For the satellites currently in orbit, we propagate under the nominal and ±2σ MSAFE forecast for Solar Cycle 26. This is captured in our reported uncertainty bounds.

### *Long Life Checks*

All remaining satellites were propagated from their initial TLE using a ballistic freefall assumption and the $C_B$ values found above, until their semi-major axis altitude fell below 280 km, as a proxy for deorbiting or falling below their operationally useful orbit. Any satellite that remained above this altitude at the end of Solar Cycle 26 (a 20 year propagation) was excluded as having a long enough life to be unaffected by the increased drag this cycle.

### *Accuracy of Ballistic Freefall Modeling for Deorbited Satellites*

The trajectories generated above were compared against actual TLE altitude trajectories, to determine if the ballistic freefall assumption accurately predicted real behavior. For satellites marked as DECAYED before 2026 in the SATCAT, the mission duration was taken as the time between the initial TLE and the time when the propagated semi-major axis altitude fell below the higher of 280 km or the final TLE altitude.

The ratio of the simulated duration versus the true duration let us assign confidence that an object was in ballistic freefall. Any ratio with an error of <10% was bucketed as HIGH confidence, and error <25% was bucketed as MEDIUM confidence. Manual spot checking of these results showed that the LOW confidence satellites displayed altitude-raising or orbit keeping maneuvers, MEDIUM confidence satellites generally appeared to be station keeping for a short period and then entered ballistic freefall (presumed to have run out of fuel), and HIGH confidence all appeared to be in ballistic freefall.

There was one exception – Planet's Dove satellites are known to be non-propulsive, and control their orbit using differential drag. This is marked by a period of higher drag at the beginning of their life for orbit acquisition and at the last month or two for rapid end-of-life deorbit. The members of FLOCK 4Y were manually marked as non-propulsive, as the initial high drag portion of flight caused them to fail this validation step, but the drag estimation picks up $C_B$ from the lower-drag operational period and is valid for future propagation. Other FLOCK groups passed this screening.

### *Accuracy of Ballistic Freefall Modeling for In-Orbit Satellites*

For satellites still on-orbit at the end of 2025 there is no observed deorbit date against which to compare, so the simulated trajectory was instead validated against the satellite's final available TLE.

We computed the cumulative dynamic pressure $Q_C$ each satellite saw up to that TLE's epoch. The ratio of simulated versus observed $Q_C$ serves the same function as the duration check. Ratios within 10% were labeled with HIGH confidence as being in ballistic freefall, and ratios within 25% were marked as MEDIUM confidence.

### First-Order Drag Calibration

The $C_B$ for the remaining objects was calibrated by scaling it by the ratio of simulated versus actual duration (or the inverse for $Q_C$ found above), to further improve its accuracy. The median $C_B$ correction for the population at this step was 1.5%, with no systemic bias. We re-ran the propagation with this corrected $C_B$ and re-screened the results with the same criteria as the last check. This was a sanity check that did not eliminate any additional satellites.

### Post-Calibration Altitude Checks for In-Orbit Objects

After $C_B$ calibration, the in-orbit objects were propagated again from their starting TLEs, to further verify their ballistic freefall assumption. Their simulated semi-major axis altitude at the time of the last TLE was compared to the observed altitude, and binned based on that discrepancy. The allowed discrepancy for HIGH confidence satellites scaled linearly from 20 km at 400 km to 5 km at 500 km, and clipped outside that range. MEDIUM confidence satellites had 3x these tolerances. Satellites whose final TLE altitude was below 350 km were exempt from this check. This scaling reflects the dynamics of decay. At low altitudes drag is strong and trajectories converge rapidly, while at higher altitudes a smaller error compounds over years of remaining life.

A satellite was only assigned an overall ballistic freefall confidence level if it passed all checks at that level.

### DISCOSweb Mass Estimates

We obtain the estimated mass of all PAYLOAD objects from DISCOSweb database and use this as a proxy for satellite costs. There were 4 satellites in our reference set launched in late 2025 where DISCOSweb did not yet have a mass populated, and these were excluded.

### Celestrak Operational Status Check

Celestrak logs operational status for on-orbit satellites, but satellites which have decayed are only marked as "D". We removed all satellites marked as non-operational. Decayed satellites with an unknown status are handled later with a knock-down factor on their economic contribution.

### Manual Operational Status Checks for High Value Satellites

For Mini-size and larger satellites (200+ kg), operational status and end-of-life cause were verified through manual research, using mission operator announcements, J. McDowell's GCAT catalog end-of-operations records, and industry reporting.18 This proved a necessary exercise, as the majority of these satellites proved to be defunct – 213 of the 303 objects examined in this category were non-operational for reasons other than drag. Causes included hardware failure, retirement, destruction, presumed abandonment (primarily Chinese EO missions), or completed short demos. The ratio of non-operative satellites was strongly correlated with age and size, especially with 60 large Soviet-era COSMOS satellites in this category.

This showed the Celestrak operational status flags to be unreliable. No satellites marked as "-" (non-operational) were found to actually be operational, but we examined more recent satellites launched since 2018 and found 12% of satellites labeled "+" (operational) were defunct. This knockdown factor is used in the cost model.

### Satellite Reference Set

Table 2 summarizes the selection and screening process performed to arrive at our reference set of 1,597 HIGH confidence operational, non-propulsive payloads.

**Table 2: Satellite Screening Process**

| Screening Step | Removed | Remaining |
|---|---|---|
| Initial Screening to LEO satellites from 2022-2026, 80-800 km | - | 13,704 |
| Megaconstellation Exclusion | 10,294 | 3,410 |
| Inert Objects, OTVs, and Human Spaceflight Exclusion | 106 | 3,304 |
| Altitude Raise Detection | 52 | 3,252 |
| VLEO Screening | 31 | 3,221 |
| Long-life Low Mass Satellites | 271 | 2,950 |
| Ballistic Coefficient Plausibility | 701 | 2,249 |
| Long Life Checks | 250 | 1,999 |
| Ballistic Freefall Validation (Deorbited) | 157 | 1,842 |
| Ballistic Freefall Validation (In-Orbit) | 50 | 1,792 |
| First-Order Drag Calibration Rescreen | 0 | 1,792 |
| Post-Calibration Altitude Checks (In-Orbit) | 32 | 1,760 |
| No Mass in DISCOSweb | 4 | 1,756 |
| Non-Operational in Celestrak | 14 | 1,742 |
| 200+ kg Non-Drag End of Life | 145 | 1,597 |

## SATELLITE COST MODEL

### *Annualized Mission Cost*

We estimate the annualized mission cost (AMC) for every satellite, where this measures a combination of the amortized capital costs and annual operational costs. Capital costs are taken as a sum of research & development effort, production costs, and launch costs. Financing, insurance, baseline expected failure rate, and the profit imperative will push the true value of a year of satellite life higher than this, so we take this as a floor above which a business would not design and build a satellite in the first place. Values are normalized to 2025 dollars.

$$AMC = \frac{R\&D + Production + Launch}{DesignLife} + AnnualOps \quad (5)$$

We value each year of lost lifetime at the full AMC under a replacement-deferral argument. The large majority of satellites in our reference set belong to serially-replenished commercial constellations, for which an additional year of on-orbit life defers the build and launch of a replacement satellite by that year – the amortized capital plus annual operations cost is therefore the avoided expenditure of a marginal year, whether that year falls within or beyond the original design life.

### *Future Value Discounting*

Nevertheless, realistic financial modeling does require discounting future value of each year of mission life. These reduce AMC for the counterfactual future lifetime $\Delta t = Life_{SWPC} - Life_{Obs}$ by a discounting scale factor, built from the decay factors $f_i$.

$$Discount = \frac{1}{\Delta t}\int_0^{\Delta t} \left(\prod 1 - f_i\right)^t dt \qquad 6$$

Firstly, we model satellite survival rates using the empirical findings from Fox et al (2013).19 Because we have already conditioned our reference set to be active satellites, we ignore infant mortality and use their steady-state exponential decay rate of 0.5% per year.

Second, we model deflation of capability-weighted forward replacement cost, assuming that economies of scale, launch costs, and capabilities per kilogram continue to improve price performance of individual satellites. Values here cover a wide range. Adilov (2022) finds that per-satellite launch costs have dropped 10.4% per year over 2000-2020, whereas Byrne and Corrado finds that bent-pipe communications satellite price-per-capacity has fallen 14.7% per year from 1963-2009.20,21 On the other hand, more bespoke missions have a slower learning rate – JWST total development cost is estimated at $9.7B against Hubble's $5.8B (1.7x, 2021 dollars), and collects light 13.6x faster rate based on Ultra Deep Field collection times.22,23,24 Over the 31-year gap between these missions, this implies a 6.6% discount rate. We model this decay as a dispersed variable with a triangle distribution centered at 10.4% with endpoints at 5% and 15%.

### *Size Class Breakdown*

The values in Table 3 below are taken as reasonable estimates for low-end AMC (educational / pathfinder missions), mid AMC (commercial satellites), and high AMC (national security, strategic civil, or flagship science missions). More detailed estimates for several constellations were performed to sanity check these ranges, including Planet's Doves, Spire's Lemurs, and an educational 3U cubesat (Nano classes), Planet's SkySat, BlackSky's, and Capella's Whitney satellites (Micro classes), and WorldView satellites (Mini Class). Note this excludes the megaconstellations, including Starlink, which benefits from SpaceX's low cost vertical integration of manufacturing and launch.

**Table 3: Satellite Annualized Mission Cost Distribution by Size Class**

| Size Class | Mass | Design Life | Low | Mid | High |
|---|---|---|---|---|---|
| Pico | 0.1-1 kg | 1-2 yr | $25k | $60k | $200k |
| Nano | 1-10 kg | 1-3 yr | $50k | $225k | $1.5M |
| Micro | 10–200 kg | 2-5 yr | $200k | $1.5M | $6M |
| Mini | 200–600 kg | 4-8 yr | $1M | $6M | $35M |
| Small | 600-1200kg | 5-12 yr | $2.5M | $15M | $60M |
| Medium+ | 1200+ kg | 7-15 yr | Bespoke estimates per-satellite | | |

### *Operational Status Knock-Down*

For satellites under 200 kg (Mini and larger were manually checked), their value is knocked down by the 12% false-positive rate measured earlier. Satellites with no definitive flag (primarily satellites with the "D" decay status but unrecorded pre-decay status) are additionally knocked down by the fraction of the reference set with definitive "+" or "-" flags (534 satellites) that are marked as non-operational (7 satellites). This levies an additional 7/534 = 1.3% reduction in value.

### *Medium+ Size Class Estimates*

Satellites over 1200 kg are expected to be high value, and their small number enables manual checking of operational status and cost estimation. These satellites in our reference set tended to have already outlived their design life, so we applied a different cost model to reflect *marginal* annualized mission cost to capture the expected value provided by each mission. The replacement-deferral argument does not hold for unique flagship missions, which are not serially replaced. For these satellites we instead construct a marginal annualized mission cost from observed lifetimes, as described below.

We find estimates for the design lifetime of these missions, and compare it to the actual observed lifetime thus far. We then designate the nominal lifetime as 1.25x the observed lifetime, and the long lifetime as the greater of 2x the design or 1.5x the observed life. The low AMC calculation is performed using low capital cost estimates with the long lifetime, the mid AMC uses mid capital costs with nominal lifetime, and the high AMC uses high capital costs with design life. This captures the range of potential mission value over bounding conditions, while centering the probability mass on the most likely scenario. The bespoke valuation for the 9 reference set satellites in this size class are listed below in Table 4.

**Table 4: Bespoke AMCs for Satellites >1200 kg**

| Satellite | ID | Design Life | Low | Mid | High |
|---|---|---|---|---|---|
| Hubble (HST) | 20580 | 15 yr | $170 M | $292 M | $967 M |
| Fermi (GLAST) | 33053 | 5 yr | $49M | $65M | $196 M |
| XRISM | 57800 | 3 yr | $62M | $108 M | $152 M |
| BlueBird (SpaceMobile -001 / -003 / -005) | 61045 61046 61047 | 5 yr | $7M | $11M | $15M |
| BlueWalker 3 | 53807 | 3 yr | $9M | $16M | $23M |
| Swift | 28485 | 7 yr | $13M | $18M | $51M |
| Einstein Probe | 58753 | 3 yr | $40M | $76M | $113 M |

We note that Fermi does have a propulsion system, but it is earmarked for collision avoidance maneuvers and end-of-life deorbiting. The others have no propulsion, reaction control thrusters, or minimally-sized fuel tanks for collision-avoidance only.

### *Monte-Carlo Approach*

AMC estimates span a full order of magnitude within the size classes, and so the low, mid, and high values are taken as the endpoints and peak of triangular probability distributions which are sampled with a 10,000 case Monte-Carlo simulation.25 The 95% confidence intervals reported here combine Solar Cycle 26's uncertainty envelope with the empirical price results.

We identified 37 satellite families covering 738 vehicles, which represent constellations from individual operators. This includes 433 of Planet's FLOCK groups and 162 of Spire's Lemur-2, with the remaining families being primarily of single-digit size. In order to capture the high correlation of satellite value within each family, we sampled the percentile draw for these satellites at the family level and applied it to all satellites in that group.

## RESULTS

### *Total Economic Impact vs 2σ High Panel Forecast*

The 1,597 satellites in our reference set lost a median of 0.70 years of mission life as compared to the 2σ high SWPC forecast, for a total of 688 years of mission life. Of these, 1,089 satellites (68%) have already deorbited, showing that much of this loss is already realized. Figure 3 shows the interesting result that since Solar Cycle 25 peaked earlier than predicted, it is falling off sooner and its strength from now until 2031 is expected to be lower than the panel predictions. This means that satellites launched since 2025 have *gained* lifetime relative to expectations, and we break out those results here.

In total, this loss of mission life equates to $0.88 billion of lost value (95% CI: $0.57B-$1.07B) for operational satellites where we have HIGH confidence in not having propulsive capabilities. Table 5 shows that this is split between $0.76B ($0.68B-$0.85B) in losses for already deorbited satellites and $0.53B ($0.34B-$0.78B) in losses for in-orbit satellites ($1.30B in total losses), with $0.41B ($0.24B-$0.73B) in gains for in-orbit satellites.

Costs are broken down by size class in Table 6 and global region in Table 7 below, for a nominal SC26. If MEDIUM confidence satellites are also included, the total impact increases by $0.20B (95% CI: $0.14B-$0.34B) for a median Solar Cycle 26. Note that a weaker SC26 means that more satellites are screened out by our criteria requiring satellites to reenter by 2041.

**Table 5: Impact By Solar Cycle 26 Behavior**

| Orbit Status | Number of Satellites | Solar Cycle 26 (MSAFE) | Median Losses | Median Gains |
|---|---|---|---|---|
| Total | 1,497 | Weak (-2σ) | $1.21B | $0.49B |
| | 1,597 | Nominal | $1.30B | $0.41B |
| | 1,629 | Strong (+2σ) | $1.21B | $0.31B |
| Already Deorbited | 1,089 | - | $0.76B | - |
| In Orbit as of 2026-06-01 | 408 | Weak (-2σ) | $0.45B | $0.49B |
| | 508 | Nominal | $0.53B | $0.41B |
| | 540 | Strong (+2σ) | $0.44B | $0.31B |

**Table 6: Impact By Size Class**

| Size Class | Mass | Number of Satellites | Median Impact |
|---|---|---|---|
| Pico | 0.1-1 kg | 229 | $11M |
| Nano | 1-10 kg | 836 | $240M |
| Micro | 10–200 kg | 506 | $429M |
| Mini | 200–600 kg | 12 | -$48M (gained) |
| Small | 600-1200kg | 5 | $25M |
| Medium+ | 1200+ kg | 9 | $220M |

**Table 7: Impact By Global Region**

| Region | Number of Satellites | Median Impact |
|---|---|---|
| United States (US) | 986 | $531M |
| Other | 291 | $163M |
| Europe (UK, FR, GER, FGER, IT, NETH, SPN, POR, GREC, BEL, LUXE, DEN, SWED, FIN, POL, CZE, CZCH, HUN, ROM, BGR, EST, LTU, LATV, SVK, SVN, HRV, ASRA, IRL, MALT, CYP, ESA, ESRO, EUTE, EUME) | 242 | $112M |
| China (PRC) | 64 | $85M |
| Russia (CIS) | 14 | -$7M (gained) |

### *Total Economic Impact vs Nominal Panel Forecast*

12 of the reference set satellites were dropped to form the nominal panel prediction case, because their propagation horizon exceeded the 2041 cutoff.

Compared to the nominal SWPC 2019 panel prediction case, a median of 1.28 years of life was lost for a total of 2,472 years across the reference set of 1,585 satellites.

This equates to $2.77 billion of lost value (95% CI: $1.96B-$3.49B) for operational satellites where we have HIGH confidence in not having propulsive capabilities. This is split between $1.37B ($1.23B-$1.53B) in losses for already deorbited satellites and $1.49B ($0.76B-$2.19B) in losses for in-orbit satellites ($2.86B in total losses), with $0.09B ($0.04B-$0.16B) in gains for in-orbit satellites. If MEDIUM confidence satellites are also included, the total impact increases by $0.50B (95% CI: $0.36B-$0.81B).

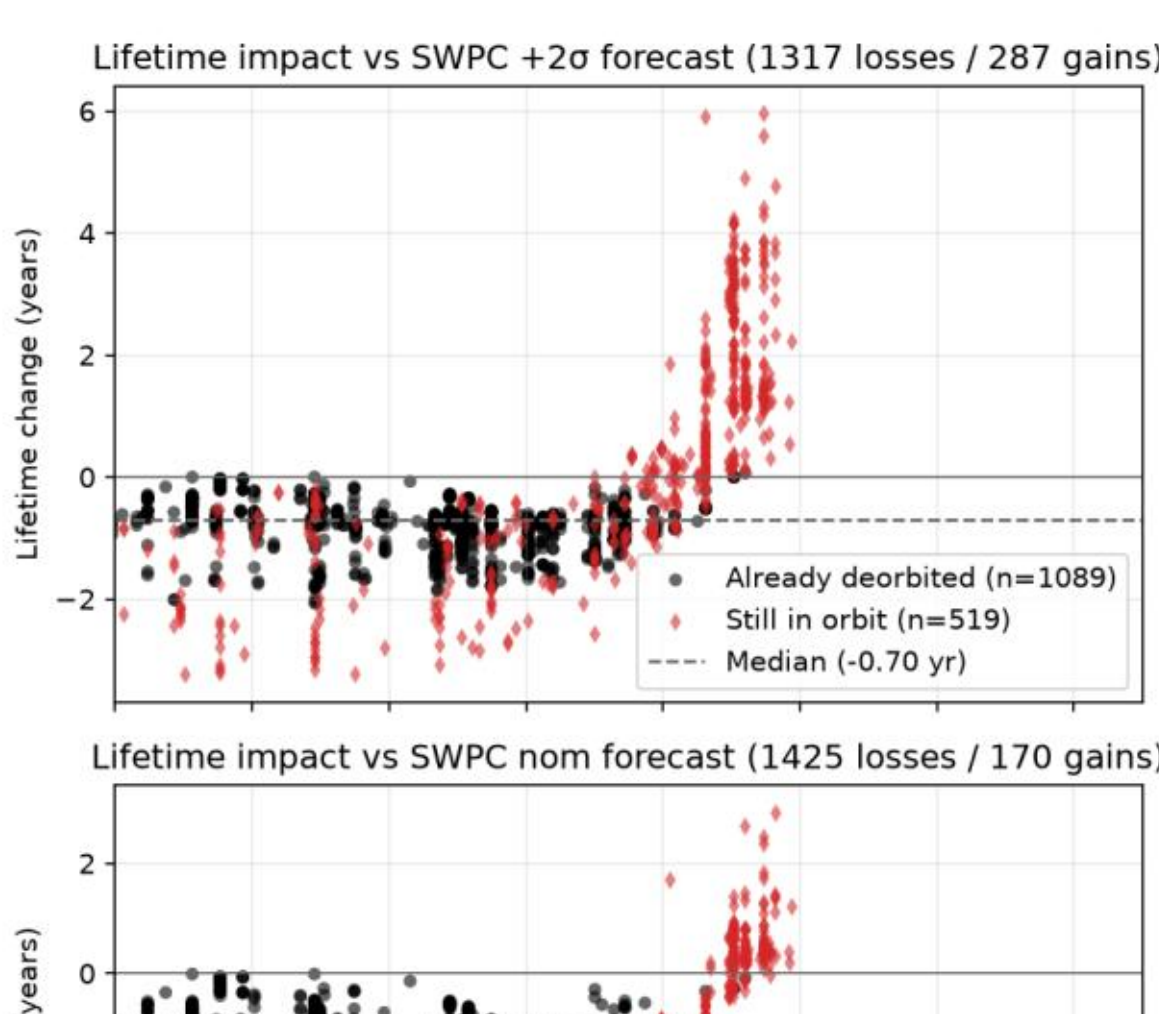

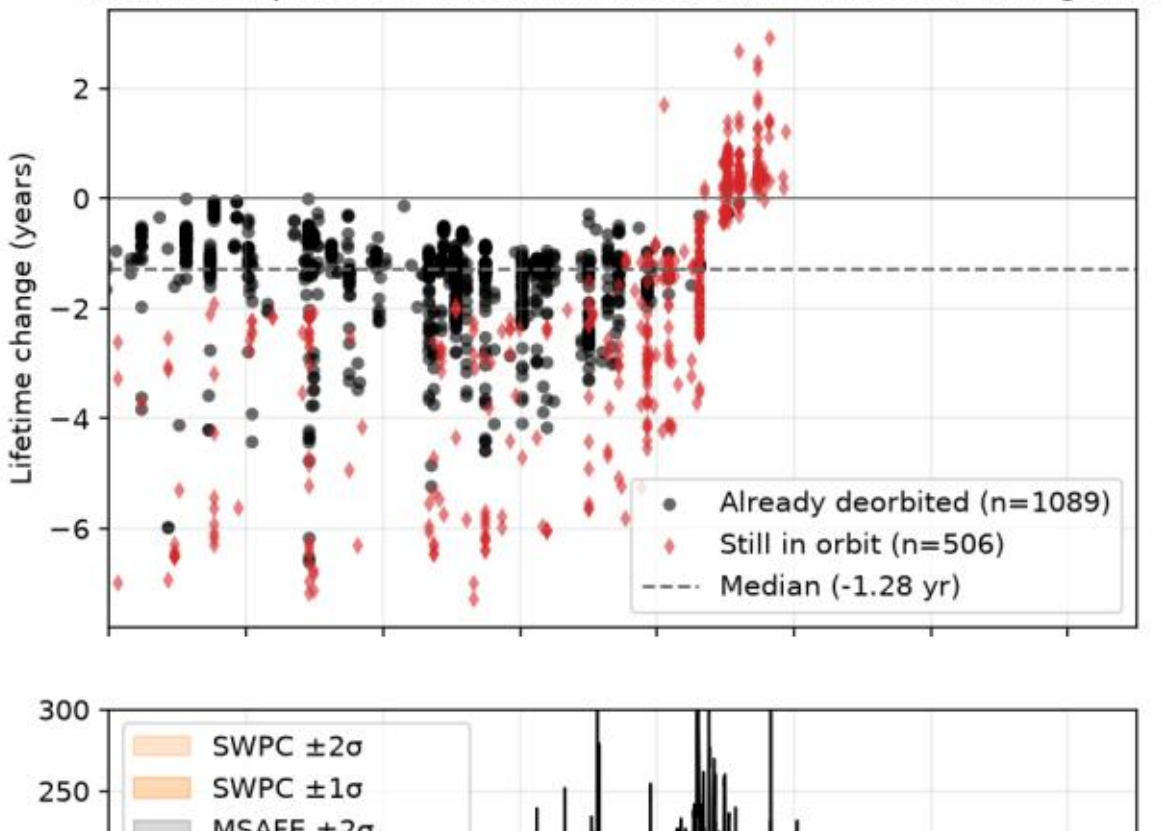

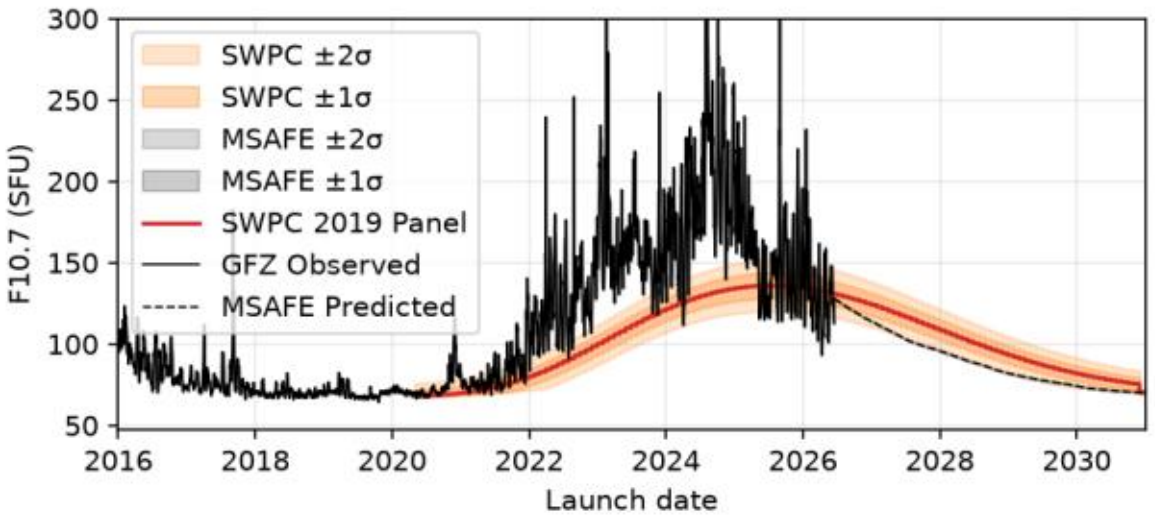

**Figure 3: Lifetime impact by satellite launch date, as compared to the SWPC panel +2σ prediction (top) and nominal prediction (middle). Solar Cycle 25 peaking sooner than expected means that solar flux (bottom) is expected to fall sooner than the panel predictions, causing satellites launched after ~2025 to gain expected lifetime relative to predictions.**

### *Hubble Space Telescope*

The Hubble Space Telescope is by far the largest single-satellite contributor to these totals, at roughly 10% of the total value lost in each case. Against the 2σ high panel prediction, it loses 0.81 years of life at $89M of value after a 0.38x discount factor. Against the nominal panel prediction, it loses 2.46 years of life at $246M of value. For Hubble this is arguably an over-conservative discount factor – using the weaker 0.5+6.6% rate for this mission we found previously, the value of this extra life would increase ~50%. See Figure 4 for the predicted altitudes used in this analysis. Note that our central estimate of 2033-2035 for Hubble's reentry aligns with NASA's latest predictions.26

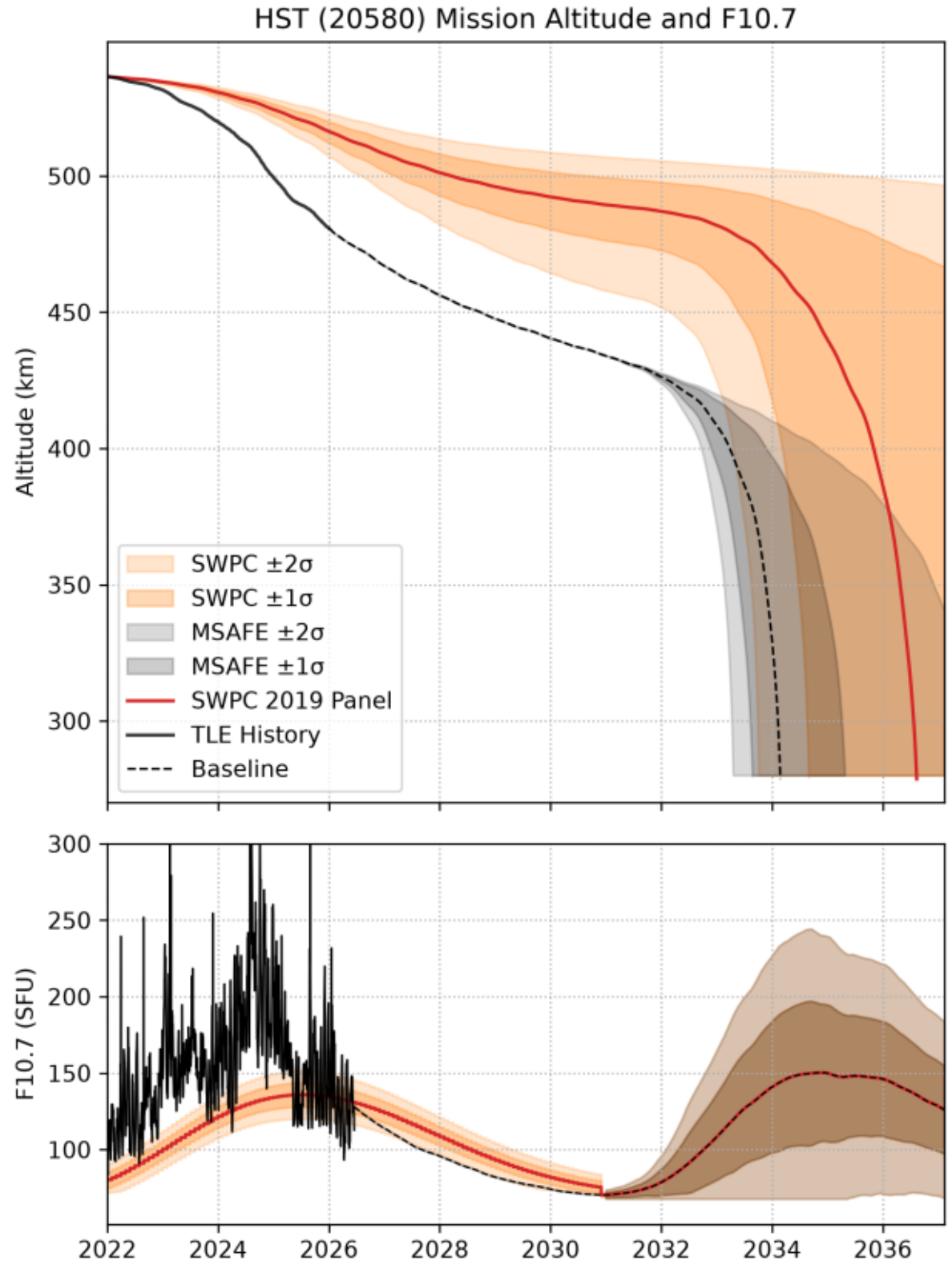

**Figure 4: Hubble's semi-major axis altitude under the different scenarios – baseline propagation based on observed and MSAFE forecasted predictions, compared to the counterfactual propagation under the SWPC forecasts. F10.7 for these scenarios is shown in the lower plot.**

### *Total Fleet Value*

Looking at all payloads in LEO at the start of 2026, our cost model reports a total median AMC annual fleet value of $33.1B/yr. This excludes crewed vehicles, space stations, and orbital transfer vehicles, zeroes inert objects, and includes custom AMC estimates for the megaconstellations and bespoke values for Medium-class and larger satellites. Operational-status knock-downs are era-dependent: satellites launched since 2011 receive the 12% flag-mismatch discount described above, while legacy satellites are discounted by measured alive-fractions (77% for those flagged operational, 6% for those with unknown status, reflecting the predominance of dead hardware among older objects). Satellites that could not otherwise be valued are assigned the median value of comparably valued satellites, scaled by the same status factors.

If we exclude the ~35% operations cost in our AMC model and assume an average 4-year mission life, then our estimate of global fleet replacement costs sums to roughly $86B.

For the 1,597 satellites in our reference set, they sum to only $1.6B/yr of AMC value and $4.1B of replacement cost (4.8% of the total). The other 95.2% of LEO fleet value lies in satellites with known or expected propulsive capabilities, or were otherwise screened out by our selection process.

Comparing to this estimated replacement cost then, Solar Cycle 25's $0.88B impact versus the 2σ high panel prediction represents 21% of the replacement value of the satellites in our reference set, and $2.77B impact versus the nominal prediction represents the loss of 68% of their value.

## DISCUSSION

### *Conservatism and Applicability to the Entire LEO Fleet*

We reemphasize that we believe this represents a conservative lower bound on the true economic impact from higher-than-expected drag during Solar Cycle 25. The 1,597 HIGH confidence satellites in our reference set represent only 12% of all payloads in LEO during this time period, and 4.8% of the total value.

Propulsive satellites were excluded from this analysis, but the 0.70 years of drag impact also applies to them. Whether the higher drag turns into actualized lifetime loss depends on fuel margins and orbit control policies on a mission-by-mission basis. While it's possible that all the satellites in the propulsive cohort contained enough extra fuel to counteract the extra $dV_{drag}$, we expect many did not. And even if there is enough fuel to meet the spacecraft's design life, this still removes the potential for economically productive mission extensions.

On the economic modeling side, the impact to downstream users will also apply an additional factor to the direct costs estimated here.

### *Sensitivity to Cycle Duration*

This analysis has also neglected variation in the length of Solar Cycle 25. If this cycle's faster onset and faster decline mean that SC26 arrives sooner than anticipated, then this will tend to increase the dollar impact of SC25's misprediction. Accounting for the whole range of solar cycle amplitude and phasing variations is critical for satellite propulsion sizing and mission planning.

### *Mitigating Drag*

Solar Cycle 25 would have behaved as it did, no matter what the SWPC 2019 panel predicted. To what extent would better predictions have mitigated this loss, as opposed to simply giving knowledge of a sunk cost?

With enough of a time horizon, satellites can be designed to handle the higher drag environment from the beginning. Closer to launch, orbit altitude can be adjusted – even only a 10-20 km change in altitude can equate to months or years of drag-induced altitude decay. After launch, operations can be changed to modify the drag profile of a satellite, or the orbit control plans for propulsive satellites can be modified to quickly boost it to less-dense altitudes, rather than maintaining altitude in a higher-density environment.

Even if the impact on any given satellite cannot be mitigated, the knowledge of the reentry timeline is critical for business operations. Realistic operational lifetime drives revenue forecasts and financing, as well as risk modeling in insurance and financial derivatives. Long-lead procurement, production, and staffing for replenishment satellites requires knowing how long the current fleet will be active, many years in advance.

### *Comparison to Risks from Radiation*

An economic impact assessment was performed by Bor et al. (2026) on the expected damages from a 1-in-100 year solar energetic particle (SEP) event.27 Over the 10,406 LEO payloads in their analysis set, they estimated $0.7B in expected damages from this sort of event, with SEP event impact concentrated in highly elliptical orbits.

Our cost models are mutually consistent. Bor et al. (2026) looks at replacement costs for operational US satellites only, and finds $43.4B in LEO fleet value. Compared to our estimate of global fleet replacement costs summing to roughly $86B, the difference is explained by the non-US fleet and our bespoke flagship mission valuation.

The estimated impact on LEO satellites from drag this past solar cycle has been about the same as the impact from this hypothetical SEP event, and solar cycles recur every 11 years rather than once a century. Moreover, the losses here have actually been realized.

### *Comparison to Risks from Orbital Debris Collisions*

Adilov et al (2023) estimated the economic impact of the orbital debris environment via a similar approach as we conducted here – expected annual direct costs as a result of collision risk.28 Over the 2,777 LEO satellites in-orbit in 2020, they estimated total fleet value of $51B-$65B, with annual expected losses of $79M-$102M (0.16%) from debris collisions. Considering the rapid growth in LEO's population since then has largely been in the lower-value megaconstellations, this compares favorably with our $86B estimate for total fleet value in 2026.

The risk numbers there come with some caveats. The analysis used ESA's Meteoroid and Space Debris Terrestrial Environment Reference (MASTER) model, which does not account for the collision avoidance capabilities of satellites avoiding other satellites or larger debris objects. On the other hand, it does not account for debris-generating events leading to a collisional cascade. Taken at face value however, a decade of the annual expected losses from orbital debris collisions matches the losses from Solar Cycle 25's misprediction.

### *Swift Case Study*

NASA's Swift observatory provides a useful case study to show the impact of high drag on operational decision making. Figure 5 shows the dramatic spread between expected and realized lifetimes. A program manager looking at a nominal expected mission life before the rise of Cycle 25 would have seen the mission lasting another decade to the start of Solar Cycle 26's rise in 2033. A realistic $2\sigma$ worst case would have given an expected deorbit in 2028. The current trajectory to deorbit in 2026 would have caught NASA as a surprise.

And in reality, this is exactly what happened. A 2019 NASA Senior Review reported a median deorbit date for Swift of 2033, with no mention of uncertainty bounds.29 It was not until April 2025 that NASA commissioned another report and discovered Swift's imminent reentry, prompting the rapid commissioning of Katalyst's reboost rescue mission.26,5

The Swift mission team ended science operations on 2026-02-11, to switch the spacecraft to a "low drag" operation mode, which we estimate increased remaining life by 8 weeks. As of writing on 2026-06-01, Katalyst's rescue mission is set to launch in June, with docking planned for July, while the spacecraft is expected to fall below its mission-specific 300 km threshold altitude in October. This is extremely tight margin on the typical timeframes of space missions. Earlier warning and continuous lifetime monitoring could have allowed detection of this issue years earlier, giving more time for low-drag flight mode changes to take effect and additional margin for Katalyst and NASA to ensure a successful rendezvous and reboost.

We estimate that if the rescue mission is unsuccessful, then Swift will have lost 2.1 years of useful lifetime against the $2\sigma$ high forecast, at $30M of value. Against the nominal case, it will have lost 6.7 years at $76M of value. But again, these are based on direct costs only. There is no other on-ground or on-orbit platform that can generate all the scientific data that Swift produces, and the downstream impact of losing this irreplaceable capability is not measured in this study.

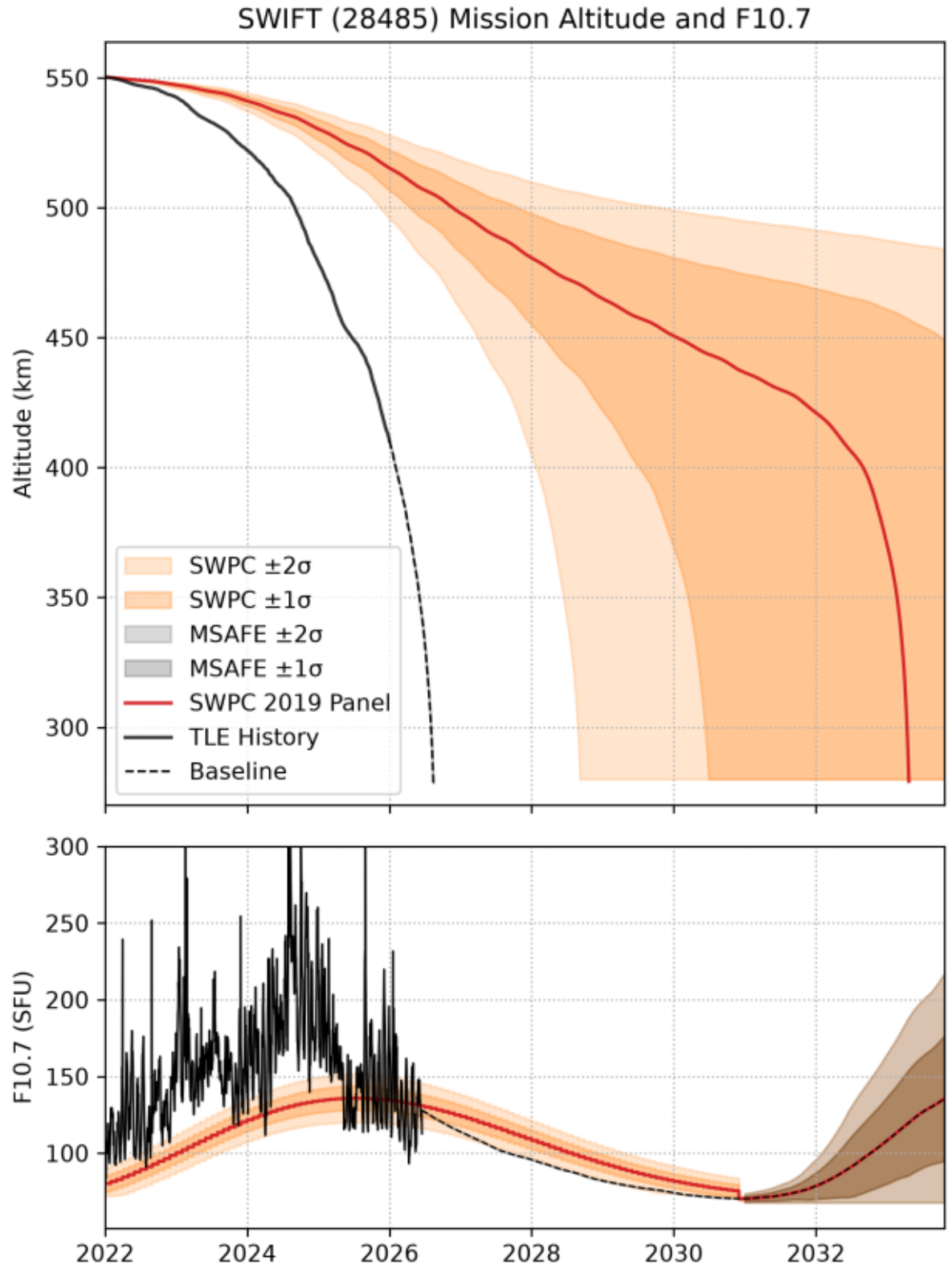

**Figure 5: Swift's semi-major axis altitude under the different scenarios – baseline propagation based on observed and MSAFE forecasted predictions, compared to the counterfactual propagation under the SWPC forecasts. F10.7 for these scenarios is shown in the lower plot.**

## CONCLUSION

Solar Cycle 25's departure from the 2019 SWPC panel forecast cost the 1,597 operational satellites in our reference set $0.88 billion in lost mission life against its two-sigma upper bound – a level that we believe a reasonable engineer would have considered safe to design to. This breaks down to $1.30B in gross losses for satellites launched before 2025, and $0.41B in gains for satellites launched since. The total impact rises to $2.77 billion in lost mission life against the nominal prediction. These are deliberate lower bounds that exclude propulsive satellites, revenue above direct cost, and downstream impact.

This realized loss rivals the expected damage from a 1-in-100-year SEP event, and it equals a full decade of expected orbital-debris collision losses. Unlike those probabilistic hazards, this one actually happened, showing decadal-scale drag uncertainty to be a leading and underappreciated space weather risk in LEO.

Two lessons follow. First, roughly two thirds of the loss ($1.89B) was avoidable by designing to the two sigma high scenario, meaning calibrated uncertainty bounds deliver value even when the central forecast misses – the remaining $0.88 billion is the cost of the bounds themselves being miscalibrated. Future prediction panels should treat uncertainty quantification as a first-class product. Second, as the Swift case shows, continuous lifetime monitoring after launch can buy operators the time to change flight modes, replan and replenish, or mount a rescue. With Solar Cycle 26 predictions now in development, Cycle 25's surprise is the clearest argument yet for investment in both forecast accuracy and calibration to protect humanity's growing presence in low Earth orbit.